# Reply to "Comments on"Stored energies and radiation Q""

Wen Geyi

It will be shown that the comment [1] contains numerous errors, misconceptions, inaccuracies, false assumptions, misunderstandings and unjustified claims. The analysis of the comment is built on a false assumption that the terminal current of an antenna fed by a waveguide is independent of frequency. This false assumption arises from the misunderstandings on the basic theory of transmission lines and misconceptions of normalization for a linear system. As a result, all the equations obtained from the assumption in the comment are incorrect, and the related discussions are untruthful. The delta gap source modeling has been abused in the comment, and this is reflected in the numerical results in the comment that give rise to negative stored energy. The main results of the commented paper have been misinterpreted by the authors of the comment, and the related analysis is established on illogical reasoning. The comment fails to pinpoint any errors in the commented paper. Instead, it reveals that the main results obtained by the authors of the comment in their previous publication [2] are incorrect.

## I. THE COMMENT IS BASED ON A FALSE ASSUMPTION THAT THE ANTENNA TERMINAL CURRENT IS INDEPENDENT OF FREQUENCY

The comment is based on the equation (1) by Yaghjian and Best [3], equation (8) by Capek et al [2], and equation (9) resulted from a combination of equation (1) and other formulations. Both equation (1) and (8), as admitted by the authors of comment, have used an assumption that the antenna terminal current is independent of frequency. Here, the terminal current refers to the modal current for the dominant mode of the feeding waveguide, as described by the equation (13) in [3]. This state of operation is illusted in Fig. 1 of the commented paper. Since the feeding waveguide is assumed to be in a single-mode operation in the neighborhood of antenna input reference plane, the transverse magnetic field in the feeding waveguide can be expressed by [4]

$$\mathbf{H}_t = I \mathbf{u}_z \times \mathbf{e}, \tag{1}$$

where $I$ is the terminal current of the antenna, $\mathbf{u}_z$ is the unit vector along the propagation axis of the feeding waveguide, and $\mathbf{e}$ is the vector modal function for the dominant mode in the feeding waveguide. The vector modal funciton $\mathbf{e}$ is independent of frequency and is determined solely by the geometry of the feeding waveguide. For example, the vector modal function for the dominant $TE_{10}$ mode for a rectangular waveguide of width $a$ and height $b$ is given by [4]

$$\mathbf{e} = -\mathbf{u}_y \sqrt{\frac{2}{ab}} \sin\frac{\pi}{a} x. \tag{2}$$

If one assumes that the terminal current $I$ is independent of frequency, the transverse magneitc field $\mathbf{H}_t$ in the feeding waveguide is also independent of frequency, leading to an awkward situation. A frequency-independent aperture field is unrealistc and would cause causality problem as pointed out by Rhodes [5]. This has also been discussed in the introduction section of the commented paper and has been ignored by the comment.

Another serious problem with the assumption is that it will introduce extra erroneous terms in the expression of the frequency derivative of reactance. In fact, we have

$$\begin{aligned}\frac{dX}{d\omega} &= \operatorname{Im}\left[\frac{dZ}{d\omega}\right] = \operatorname{Im}\left[\frac{d}{d\omega}\frac{V(\omega)}{I(\omega)}\right] \\ &= \operatorname{Im}\left\{\frac{1}{I(\omega)}\frac{dV(\omega)}{d\omega} - \frac{Z(\omega)}{I(\omega)}\frac{dI(\omega)}{d\omega}\right\},\end{aligned} \tag{3}$$

where $Z = R + jX$ denote the antenna input impedance, $V$ is the terminal voltage. If the the terminal current is assumed to be independent of frequqncy, the above equation reduces to

$$\frac{dX}{d\omega} = \operatorname{Im}\left\{\frac{1}{I(\omega)}\frac{dV(\omega)}{d\omega}\right\}. \tag{4}$$

Hence there is a startling difference between (3) and (4) due to the incorrect assumption. This explains why the equations (1), (8) and (9) in [1] have extra terms compared with the one without using the incorrect assumption. These extra terms are erronous whereas they have been claimed by the authors of the comment as a valuable finding.

As a matter of fact, all the field and circuit quantities in microwave frequency regime depend on frequency, and this is a well-known fact. Also note that, in practice, the terminal voltage and current essentially respresent the Fourier transforms of their time-domain counterparts and they thus depend on frequency. Since both the comment and [2] make use of the incorrect assumption as the starting point, all the related equations obtained from the assmuption, especially (1), (8) and (9) in [1] and the main equations in [2], are invalid for an antenna with feeding waveguide connected, and thus all the discussions associated with these equations are false.

## II. NORMALIZATION FOR A LINEAR SYSTEM IS ABUSED IN THE COMMENT

The comment asserts that the terminal current can be assumed to be independent of frequency by "normalization" since the antenna system is linear (See the footnote of [1]). This statement is misleading, and actually incorrect for an antenna system. For a linear system characterized by a linear operator $L$, the output $y$ is related to the input $x$ (a known quantity) by

$$y = L(x). \tag{5}$$

For any constant $c$, we have the following linearity

$$cy = L(cx). \tag{6}$$

The constant $c$ can be arbitrarily chosen so that the input $x$ or the output $y$ has a specific magnitude, and this process is called normalization. Usually both the input and output depend on some parameters (e.g, space position, time or frequency). The normalization process should not change the functional relationship of the input or output with these parameters. In other words, the graph of the input or the output with respect to these parameters should maintain the same shape after the normalization except for a proportionality constant.

For an antenna system connected to a feeding waveguide, the input (the known quantity) is the incident field from the feeding waveguide. Both the terminal voltage $V$ and current $I$ are an unknown quantity to be determined and cannot be arbitrarily specified or normalized. To illustrate this, let us consider a typical antenna problem: an aperture antenna fed by a waveguide as shown in Fig. 1. Assume that the feeding waveguide is in a single-mode operation. The incident field coming from the left is then given by (a known quantity)[4]

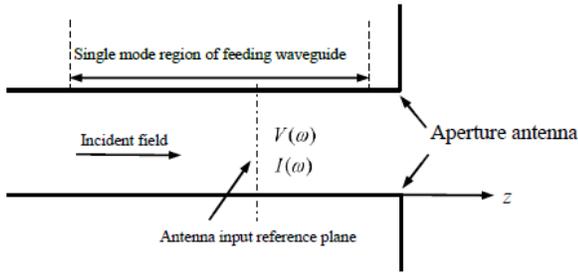

Fig. 1 Aperture antenna fed by a waveguide.

$$\mathbf{E}_{in} = ce^{-j\beta z}\mathbf{e}, \quad (7)$$

where $\mathbf{e}$ is the vector modal function for the dominant mode, $\beta = \sqrt{k^2 - k_{c0}^2}$ is the propagation constant with $k_{c0}$ being the cut-off wavenumber for the dominant mode, and $c$ is an arbitrary constant and can be chosen (or normalzied) to be unit. When the incident field encounters the aperture (the discontinuity), a number of higher order modes in the feeding line will be excited in the neighborhood of the discontinuity. Most of the energy will be radiated into free space and part of the energy will be reflected back to the feeding line. At the antenna input plane (reference plane), where only the dominant mode is assumed to be propagating, the terminal voltage and current can be expressed as[4]

$$V(\omega) = c[e^{-jkz} + \Gamma(\omega)e^{jkz}],$$
$$I(\omega) = c[e^{-jkz} - \Gamma(\omega)e^{jkz}]/Z_w(\omega), \quad (8)$$

where $\Gamma(\omega)$ and $Z_w(\omega)$ are the reflection coefficient and the wave impedance for the dominant mode respectively. In order to determine the terminal voltage $V(\omega)$ and current $I(\omega)$, we need first to solve a boundary value problem to figure out the reflection coefficient $\Gamma(\omega)$. The terminal voltage and current clearly depend on the frequency and cannot be arbitrarily specified or normalized since they both contain an unkown function $\Gamma(\omega)$ and they are the derived quantities(Indeed, if the terminal current could be assumed to be independent of frequency, other field quantity such as the electric field generated by an antenna might also be assumed to be independent of frquency if one follows the same line of reasoning as the comment). The only quantity that can be adjusted is the amplitude $c$ of the incident field, which can be normalized to unit as usual.

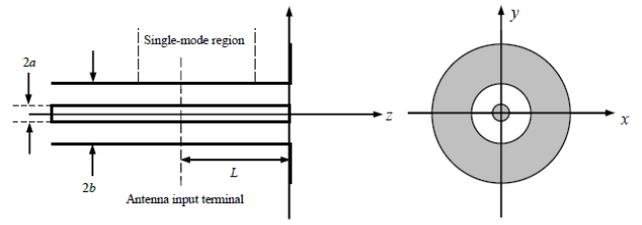

Fig. 2 A coaxial aperture with infinte flange.

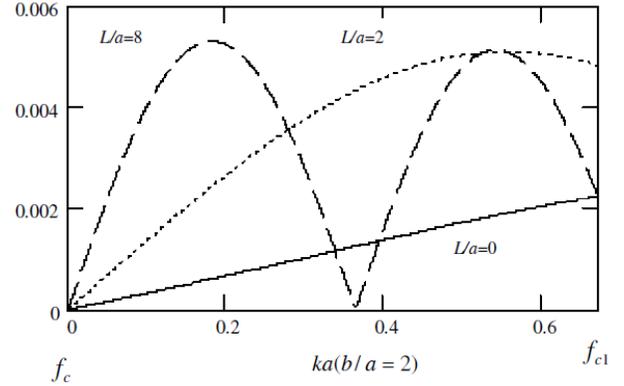

Fig. 3 The modal current $|I|$ of the coaxial aperture antenna with infinite flange.

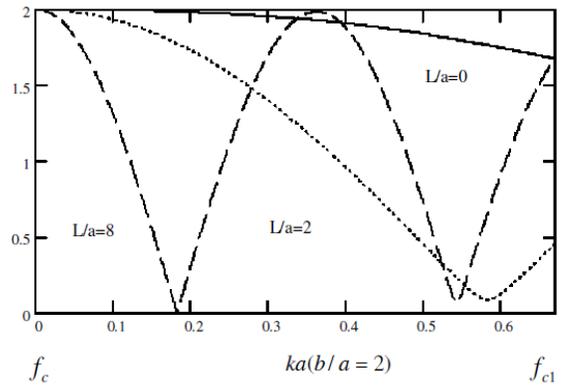

Fig. 4 The modal voltage $|V|$ of the coaxial aperture antenna with infinite flange.

Fig. 2 shows a coaxial aperture antenna with an infinte flange. The aperture antenna is excited by the incident dominant TEM mode with $c = 1$, and has been studied by a rigorous integral equation approach in [6]. Fig. 3 and Fig. 4 respectively show how the terminal current and voltage for a coaxial aperture antenna vary with the frequency as well as the location of the input terminal. Note that all the calculations are limited to the frequency range from the cut-off frequency of the dominant mode ( denoted $f_c$ ) to that of the first higher order mode ( denoted $f_{c1}$ ). This guarantees a single-mode operation in the feeding waveguide. Note that the terminal voltage and current vary rapidly with frequency over the whole frequency range, clearly indicating that it is generally wrong to assume a frequency-independent voltage or current in the feeding waveguide.

III. THE RESULTS OF THE COMMENTED PAPER HAVE BEEN MISINTERPRETED AND THE RELATED ANALYSIS OF THE COMMENT IS BASED ON ILLOGICAL REASONING

Equations (4) and (5) in the commented paper are the Foster reactance theorem for an ideal antenna and are derived in a previous publication [7], which relies on the assumptions that the antenna has no heat loss and is connected to a feeding waveguide in a single-mode operation, while the equations (29), (30) and (31) in the commented paper are new results derived from the conditions that the antenna is characterized by a current distribution and no feeding waveguide and thus no terminal current is involved. The comment has misinterpreted these results and has falsely claimed that these new results should also be valid for the frequency-independent terminal current (See the footnote of the comment). In fact, a frequency-independent terminal current that leads to a frequency-independent transverse magnetic field at the feeding aperture would induce an unrealistic current distribution on the antenna surface. Once this unrealistic current distribution is introduced into (29), (30) and (31), erroneous results may occur.

Furthermore, the authors of the comment, on the one hand, assert that the Foster reactance theorem given by (4) and (5) in the commented paper is incorrect, and on the other hand, they combine the equations (4) and (5) and the new equation (29) in the commented paper to obtain (6) in the comment to evaluate the frequency derivative of reactance of a strip dipole antenna of finite width. The numerical results obtained this way agree very well with those from FEKO simulation as well as by equation (6) in the low frequency range but disagree around the anti-resonances (where negative slope of reactance occurs) in the high frequency range, as indicated by Figs.1 and 2. From these results, the authors of the comment made a hasty conclusion that the new expression (29) in the commented paper is incorrect. The above reasoning is illogical. According to the basic concepts of logic, a false statement C (e.g., the disagreement of the numerical results in high frequency range) resulted from a false statement A (e.g., the 'incorrect' Foster reactance theorem as assumed by the comment) and a statement B (e.g., the new expression (29)) does not imply that the statement B is false.

Now we will illustrate that the numerical results depicted in Figs. 1 and 2 are problematic.

IV. THE DELTA GAP SOURCE MODELING IS ABUSED IN THE COMMENT WHICH CAUSES NEGATIVE STORED ENERGY

The modeling of wire antennas usually depend on a number of approximations [8]-[16]. The most dramatic approximation is that the feeding line is replaced by a delta gap (also known as delta function generator) or magnetic frill generator [8]. Physically the delta function generator represents a point source and is a pure mathematical model that simplifies the excitation region of antenna by assuming that the incident electric field from the feeding line exists only in the gap between the two wire terminals of the antenna and is zero outside. When the delta gap is used to model a wire antenna, either the terminal voltage or the current is considered as the input (a known quantity instead of a derived quantity) and thus can be normalized. This assumption is essentially a low frequency approximation and has been widely adopted by various simulation tools such as the FEKO being used by the comment. The delta gap source modeling is questionable, and cannot be checked experimentally as discussed by Wu, King and Schmitt [11]-[12] since all practical antennas involves a feeding waveguide, and the feeding waveguide itself contributes significantly to the value of antenna impedance [9]. Maloney, Smith and Scott have also discussed this [13] and the following is a quote from their paper:

"The theoretical model used for the antenna usually involves approximations introduced to simplify the analysis. For example, for the cylindrical dipole antenna an idealized source is often used, the so-called "delta-function generator". This source does not correspond to any realizable experimental model. Often the equations involved in the analysis of the antenna are also approximate. For example, for the cylindrical dipole antenna approximate integral equation (thin wire approximation) is often substituted for the exact integral equation. Approximations, like those mentioned above, lead to discrepancies between theoretical and experimental results, and it is often difficult to quantitatively ascertain the effects of the different approximations."

Also as pointed out by Balanis [8], the delta-gap source modeling is the simplest and most widely used, but it is also the least accurate, especially for antenna impedances although it generally performs well for radiation patterns. Some efforts have been devoted recently to improve the accuracy of the delta gap source modeling in numerical methods [14]-[16]. From mathematical point of view, the delta function generator is exact only if the wire antenna is infinitely thin and the gap is infinitely small. When the radius of the wire is finite, the delta function generator is accurate for the impedance only in the low frequency range. This fact has been overlooked by some researchers including the authors of the comment.

The equation (1) obtained by Yaghjian and Best and the equation (8) obtained by Capeck and Jenlinek in [1] use the same assumption as the delta function generator does, i.e., the terminal current at the feeding point is assumed to be independent of frequency. Therefore equations (1), (8) and (9) may be considered as the results derived from the delta gap source modeling, but they cannot be applied to an antenna connected to a feeding waveguide as discussed before.

The comment has demonstrated some numerical results of the frequency derivative of reactance for a strip dipole of finite width and a Yagi-Uda array based on the delta gap source modeling (Fig. 1 and Fig. 2 of [1]). It is shown that the results from equations (8) and (9) in [1] agree very well with FEKO simulation. This is not surprising since the equations (8) and (9) in [1] and the FEKO simulation tool use the same delta gap source modeling (Also note that the current distributions used in the equations (6), (8) and (9) are from the FEKO simulation with a delta gap excitation). However the authors of the comment fail to note that their numerical results shown in Figs. 1 and 2 are accurate for the impedance only in the low frequency range since the strip dipole has a finite width. In fact, the frequency derivative of reactances depicted in Figs. 1 and 2 in [1] stand for the total stored energy of the antenna according to (6), (8) and (9) in [1]. It can be seen from Figs. 1 and 2 that the total stored energy can be negative in the high frequency range, which is physically unacceptable. It is quite strange that the comment has removed the frequency derivatives of reactance in the low frequency range (which is more accurate) and only exhibited the frequency derivatives of reactance in the high frequency range (which is less accurate) in Fig. 2.

We finally note that the last two terms in (6) and (8) in [1] are insignificant in the low frequency range (See the discussion in the commented paper), which is the reason why the numerical results from (6) and (8) in [1] agree in the low frequency range.

## V. FURTHER NOTES ON FOSTER REACTANCE THEOREM FOR AN IDEAL ANTENNA

The comment asserts that Foster reactance theorem for an ideal antenna (i.e., without heat loss) described by (4) and (5) in commented paper is incorrect. They have used some references (references 2 and 3 in [1]) to support their assertion and totally disregarded the discussions in the commented paper as well as in the related references [17][18], which are missing in the reference section of the comment. We stress again that the Foster reactance theorem has been derived for an ideal antenna connected with a feeding waveguide in a single-mode operation. Although the feeding waveguide is assumed in the derivation of the Foster reactance theorem, we find, from numerous numerical simulations[6][17][18], that the Foster reactance theorem also applies for an infinitely thin wire antenna excited by a delta function generator.

The Foster reactance theorem for antenna has been a controversial topic for many years. The controversy comes from a number of numerical simulations (e.g., Figs. 1 and 2 in the comment), indicating that negative frequency derivatives of reactance may occur around anti-resonances in the high frequency range, contradicting that the frequency derivatives of reactance must be positive for an ideal antenna as predicted by Foster reactance theorem. It must be mentioned that majority of the numerical simulations come from the delta gap approximation, including the reference 3 cited in [1]. Strictly speaking, the controversy originated from the delta gap simulations can be ignored since the Foster reactance theorem for an ideal antenna is only rigorously proved for the ideal antenna with a feeding waveguide connected. Some researchers, including the authors of the comment, fail to note that the delta gap approximation is accurate for the impedance only in the low frequency range (below the first anti-resonant frequency) for a wire antenna of finite radius, and that the negative frequency derivatives of reactance will gradually disappear when the radius of the wire antenna approaches to zero (The delta function generator approximation becomes exact in this case). This has been illustrated in Fig. 5, where the reactances of a center-fed dipole antenna of length 0.15m with different radius are depicted.

It is worth mentioning that a rigorous method for the analysis of an antenna with a feeding waveguide connected has been proposed by the author in [6]. The method is based on a combination of the integral equation with the field expansions in the feeding waveguide. It has the advantage of high accuracy in antenna input impedance calculation as it does not involve any approximations in the antenna source region and thus is more realistic and accurate. Another advantage of introducing the feeding line in the integral equation formulation is that it guarantees a unique solution, thus providing a solid basis for the analysis of various antennas. Some typical antennas are analyzed in [6] and the numerical results indicate that the Foster reactance theorem holds for all antennas investigated.

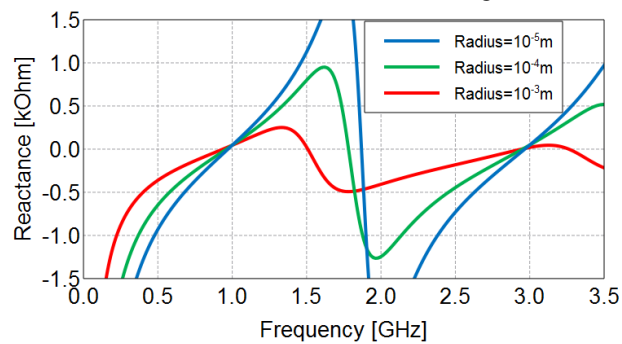

Fig. 5 Reactances of a dipole antenna with different radius from FEKO simulations using delta gap source modeling.

## VI. THE COMMENT IS CRAMMED WITH UNJUSTIFIED CLAIMS

The authors of the comment have made several unjustified claims such as the results obtained in their previous publication [2]. Especially they have stressed that the commented paper did not mention their publication [2]. It should be noted that the commented paper and the reference [2] deal with two different antenna models. The former investigates the antenna model characterized by a current distribution without a feeding mechanism while the latter studies the antenna with a feeding structure where the antenna terminal is defined. It must also be pointed out that the commented paper was submitted to TAP on March 20, 2013 while the paper [2] was submitted to TAP on May 09, 2013. Considering the fact that the main results in [2] are invalid for an antenna connected with a feeding line, this becomes an irrelevant issue.

## VII. CONCLUSION

In summary, the comment by Capek and Jelinek is based on equations (1), (8) and (9) in [1], which are established on an incorrect assumption that the terminal current of an antenna connected to a feeding waveguide is independent of frequency. For this reason, the equations (1), (8) and (9) in [1] are not applicable to any practical antenna systems. The comment contains misunderstandings about the basic theory of transmission lines, misconception of normalization for a linear system, inappropriate numerical results originated from the delta gap source modeling (which lead to negative stored energy), illogical reasoning and unjustified claims. In contrast, the approach in the commented paper is straightforward, rigorous, and does not require any redundant assumptions. The comment fails to identify any mistakes in the commented paper and is thus superfluous.